# "Clumpiness" Mixing in Complex Networks

Ernesto Estrada,[1*] Naomichi Hatano,[2] Amauri Gutierrez[3]

[1]*Complex Systems Research Group*, X-rays Unit, RIAIDT, Edificio CACTUS, University of Santiago de Compostela, 15706 Santiago de Compostela, Spain

[2]Institute of Industrial Science, University of Tokyo, Komaba 4-6-1, Meguro, Tokyo 153-8505, Japan

[3]Bayes Inference S.A., Gran Vía 39, 5ta Planta, 28013 Madrid, Spain

---

[*] Corresponding author. E-mail: estrada66@yahoo.com. Fax:+34-981-547-077.

**Abstract.** Three measures of *clumpiness* of complex networks are introduced. The measures quantify how most central nodes of a network are clumped together. The assortativity coefficient defined in a previous study measures a similar characteristics but accounts only for the clumpiness of the central nodes that are directly connected to each other. The clumpiness coefficient defined in the present paper also takes into account the cases where central nodes are separated by few links. The definition is based on the node degrees and the distances between pairs of nodes. The clumpiness coefficient together with the assortativity coefficient can define four classes of networks. Numerical calculations demonstrate that the classification scheme successfully categorize 30 real-world networks into the four classes of clumped assortative, clumped disassortative, loose assortative and loose disassortative networks. The clumpiness coefficient also differentiates the Erdös-Rényi model from the Barabási-Albert model, which the assortativity coefficient could not differentiate. In addition, the bounds of the clumpiness coefficient as well as the relations among the three measures of clumpiness are discussed.

# 1. Introduction

Graph and networks are ubiquitous in physics, ranging from fundamental physics [1-4] to applied socio- and econophysics [5, 6]. Despite the long tradition in studying physical objects by means of graphs, novel applications continuously arise posting new challenges for theoretical and mathematical physicist. The recent explosion of works studying "complex networks" is one of the sources of new concepts and theoretical problems [7-11]. The best-known examples are the concepts of "small-worldness" [12] and "scale-freeness" [13], which have produced an avalanche of new results in this field [7-11]. Another area of intensive research is stimulated by the necessity of defining new measures characterizing the topological structure of complex networks [14], such as the identification of the most central nodes in a complex network [15]. These studies on network "centrality" are crucial for understanding several effects on complex networks. Among them, we can mention the resilience of networks to intentional attacks [16], the identification of the most influential individuals in a social network [17] as well as the protection of the keystone species in an ecosystem [18, 19]. By *central*, we mean a node having the largest value of a graph theoretic parameter (centrality) [20], which characterize a topological property of this node in the network, such as its number of connections (degree) [17], the number of shortest paths passing through it (betweenness or load) [17], its relative closeness to the rest of nodes in the graph (closeness) [17], or its participation in all substructures of the network (subgraph centrality) [21].

It has been shown that the identification of the most central nodes in a network is not enough for solving several practical issues. For instance, it has been found that if the most central nodes are clumped together in a network, the consequence for network resilience, transmission of an epidemics or ecological conservation are quite different from the cases where they are spread across the network [22]. Newman introduced an "assortativity" coefficient as a measure to quantify this characteristic of certain complex networks [22]. This index is simply the Pearson correlation coefficient of the degrees at either ends of a link. It thus accounts only for the "clumpiness" of the central nodes that are directly connected to each other in the network. In the cases where the central nodes are separated by two or only few links, the network can display disassortative properties in spite of the fact that the most central nodes are practically clumped together in the graph.

We here propose measures that account for the "clumpiness" of the most central nodes in a network. The measures defined here are referred to the *clumpiness* coefficient (Section 3) and the spectral measure of clumpiness (Section 4). A desired characteristic of the measures is that they have the maximal value when the most central nodes are as close as possible. The clumpiness should decrease when we reduce the centrality of the nodes. In addition, the increment in separation of these central nodes should also decrease the clumpiness of the network. We then present in Section 5 numerical calculations of the clumpiness measures of various networks. In particular, we propose categorizing networks on the basis of combination of the clumpiness and the assortativity.

## 2. Preliminaries

### 2.1. Elementary definitions

Before going into the study of the clumpiness, let us first present some elementary definitions as well as state our motivation of defining the clumpiness. A *graph invariant* is defined to characterize an inherently graph-theoretic *property* of a graph [23]. It is defined as a measure based on *graph parameters* that do not change with a change of the labels of nodes/links. By graph parameters, we understand any local or global topological property of a graph, such as node/link properties, matrix or vector representation of the graph, *etc.* Here we are dealing with simple, connected graphs $G = (V, E)$, where $V$ is the set of nodes of cardinality $|V| = n$ and $E$ is the set of links representing relationships between the nodes. The degree of a node $i$, also known as the degree centrality, is designated as $k_i$ and it is equal to the number of links incident to $i$. The topological distance $d_{ij}$ is the minimum number of links separating the node $i$ from a node $j$ [24]. Vectors and matrices will be represented by lower case and upper case bold letters, respectively.

Those graphs that can be transformed to each other by simply changing the labeling of the nodes are called *isomorphic*. More formally, two graphs $G = (V, E)$ and $G' = (V', E')$ are *isomorphic* if there exists a one-to-one function, called an *isomorphism*, from $V$ onto $V'$ such that $pq \in E$ if and only if $p'q' \in E'$ [25]. Any graph invariant is exactly the same for any pair of isomorphic graphs. However, there are pairs of nonisomorphic graphs that have identical values of certain graph invariants. These graphs will be called here to be *degenerate* with respect to this graph invariant. The *discriminant power* of a graph invariant is simply the proportion of nonisomorphic graphs which are differentiated by a graph invariant.

**2.2. Motivation**

The topological structure of complex networks is also complex. Consequently, the architectural organization of complex networks is not expected to be characterized by a single index or measure. A typical example of this situation is the characterization of a network on the basis of its node degrees. A now "classical" way of such characterization is to use the degree distribution, which indicates the probability of finding a node of certain degree (or range of degrees) in the network. Accordingly, a network can display a uniform, exponential or power-law degree distribution of its node degrees. The degree distribution, however, tells us nothing about the correlation between nodes. For instance, if a network has a power-law degree distribution, we know that there is a low probability of finding a high-degree node in the network, but nothing is said about whether the high-degree node is connected to another high-degree node or to a low-degree one. Consequently, we can consider the degree distribution as a "zeroth-order" measure or index of a complex network.

A step forward in the characterization of the organization of nodes in a network is to measure how the nodes are connected to each other. The assortativity coefficient is a naïve characterization of this situation, in which we obtain information as to how high-degree and low-degree nodes are connected to each other. A positive assortativity coefficient indicates that high-degree nodes are preferentially attached to other high-degree nodes. On the other hand, a negative assortativity coefficient indicates that high-degree nodes are preferentially connected to low-degree nodes. Consequently, the assortativity coefficient is a "first-order" measure or index of a complex network.

A first-order measure such as the assortativity coefficient tells us nothing about the way in which nodes are organized beyond the nearest neighbors. For instance, in an assortative network, some high-degree nodes are linked to other high-degree nodes, but some high-degree nodes can be separated by very few links or by long paths. In the former case, the high-degree nodes form a clumped cluster while in the latter they are spread across the network. Neither of these two situations are distinguished by the assortativity coefficient as it attempts to characterize only the "first-order" topological characteristics of the network.

A real-world example of this situation is illustrated in Figure 1. The network illustrated in Figure 1A corresponds to the inmates in a prison and that in Figure 1B to the food web of St. Marks. Both networks are almost of the same size, $N = 67$ and $N = 48$, respectively, both display uniform degree distributions and have almost identical assortativity coefficient, $r = 0.103$ and $r = 0.118$, respectively. However, while in the prison network the high-degree nodes are spread across the network, they are clumped together in the food web. This difference can have dramatic implications for the structure and functioning of these two systems.

**Insert Figure 1 about here.**

In a similar way we can find disassortative networks, where high-degree nodes are preferentially attached to low-degree nodes, we can also find that the high-degree nodes can be separated by only two links with a low-degree node acting as a bridge or by very long paths. This situation is illustrated in Figure 2 for a sexual network in Colorado Springs (A) and the transcription interaction network of *E. coli* (B), which have almost equal negative assortative coefficients. In the former case the high-degree nodes are separated by very long chains while in the latter case most of the high-degree nodes are clumped together separated by only two or three links.

**Insert Figure 2 about here.**

## 3. "Clumpiness" coefficient

### 3.1. The definition of the clumpiness coefficient

The clumpiness coefficient of the degree centrality in the graph $G$ is defined here by the expression

$$\Lambda(G, k, \alpha) = \sum_{i>j}^{n(n-1)/2} \frac{k_i k_j}{(d_{ij})^\alpha}, \qquad (1)$$

where $\alpha > 0$ is a real parameter. Our motivation for using an inverse power-law potential in expression (1) is because of its similarity with several well-known potentials, such as the Coulombic and gravitational ones, as well as others accounting for the inter-molecular interactions, e.g., Lennard-Jones potential. According to the above definition, the clumpiness coefficient increases with the increase of the degrees of the nodes in the network and decreases with the increase in the separation between these nodes.

As the selection of the most appropriate value for $\alpha$ here remains empirical we have calculated the clumpiness coefficient $\Lambda(G,k,\alpha)$ for different values of this parameter, namely $\alpha = 1,2,3,4,5$ for the series of 19 cubic regular graphs. These graphs have 10 nodes and 15 links with all nodes having degree equal to 3. For each value of $\alpha$ we normalized the values of $\Lambda(G,k,\alpha)$ by the maximum value of the clumpiness obtained for this series of graphs. Then, we have plotted the values of the normalized clumpiness coefficient for the 19 graphs for every specific value of $\alpha$. As can be seen in Fig. 3 the clumpiness coefficient for $\alpha = 1$ have different values for 11 of the 19 graphs. The rest of the clumpiness coefficients for $\alpha = 2,3,4,5$ discriminate 12 out of the 19 graphs. In addition, the largest percentage of variation in the normalized clumpiness coefficient also displayed in Fig. 3 is obtained for $\alpha = 2$. Consequently, we select the value of $\alpha = 2$ in the clumpiness coefficient (1) for the rest of the calculations to be carried out in this work. It is worth mentioning that the results obtained with $\alpha = 2$ are very similar to those obtained for $\alpha = 1$, which corresponds to the well-known Coulombic and gravitational potentials.

**Insert Figure 3 about here.**

Let $\mathbf{k} = \begin{pmatrix} k_1 & k_2 & \cdots & k_n \end{pmatrix}$ be a row vector of the node degrees in the graph and let $\mathbf{D}$ be the topological distance matrix of the graph, whose $(i,j)$ element is $d_{ij}$. We will denote by the symbol $\mathbf{D}^{*\alpha}$ the $\alpha$ th entrywise power [26] of $\mathbf{D}$, that is, a matrix in which every entry of $\mathbf{D}$ is raised to the power $\alpha$. Then, let $\mathbf{R}$ be the matrix defined as

$$\mathbf{R} = (\mathbf{D}+\mathbf{I})^{*-2} - \mathbf{I}, \qquad (2)$$

where $\mathbf{I}$ is the identity matrix. $\mathbf{R}$ is the matrix whose elements are given as follows:

$$R_{ij} = \begin{cases} (d_{ij})^{-2} & \text{for } i \neq j, \\ 0 & \text{for } i = j. \end{cases} \qquad (3)$$

Then, we have

$$\Lambda(G) = \frac{1}{2}(\mathbf{k}^T \mathbf{R} \mathbf{k}). \tag{4}$$

The formula (4) was originally proposed by Estrada *et al.* 10 years ago when studying modifications of the Harary-like topological indices in chemistry [27].

In a similar way, the clumpiness coefficient can be obtained from a clumpiness matrix, which is defined as follows. Let $\mathbf{K} = diag(k_1, k_2, \cdots, k_n)$ be a diagonal matrix of the node degrees of the graph and let $\mathbf{R}$ be the matrix previously defined. Then the clumpiness matrix is

$$\hat{\mathbf{I}} = \mathbf{KRK}, \tag{5}$$

whose $(i, j)$-entries are $\dfrac{k_i k_j}{(d_{ij})^2}$ and the diagonal entries are zeroes. The clumpiness matrix will find other important applications in the current work. The clumpiness coefficient is then obtained as the half-sum of the entries of this matrix

$$\Lambda(G) = \frac{1}{2}(\mathbf{u}^T \hat{\mathbf{I}} \mathbf{u}), \tag{6}$$

where the vector $\mathbf{u}$ is an all-one vector.

### 3.2. Clumpiness coefficient for certain classes of graphs

We now calculate the clumpiness coefficient explicitly for four classes of graphs. Let $P_n$, $C_n$, $S_n$ and $K_n$ be the path, cycle, star and complete graphs of $n$ nodes, respectively [28]. We obtained the following formulas for the clumpiness coefficient in such graphs:

$$\Lambda(P_n) = \frac{1}{(n-1)^2} + 4\sum_{x=1}^{n-2} \frac{1}{x^2} + \sum_{x=1}^{n-3} \frac{4(n-x-2)}{x^2}, \tag{7}$$

$$\Lambda(C_n) = \begin{cases} 8/n + 4n \sum_{x=1}^{n/2-1} \dfrac{1}{x^2} & \text{for } n \text{ even}, \\ 4n \sum_{x=1}^{n/2-1} \dfrac{1}{x^2} & \text{for } n \text{ odd}, \end{cases} \tag{8}$$

$$\Lambda(S_n) = \frac{(n-1)(9n-10)}{8}, \tag{9}$$

$$\Lambda(K_n) = \frac{n(n-1)^3}{2}. \tag{10}$$

For large values of $n$ we have

$$\Lambda(P_n) \approx 4\sum_{x=1}^{n-2}\frac{1}{x^2} + 4(n-2)\sum_{x=1}^{n-3}\frac{1}{x^2} - 4\sum_{x=1}^{n-3}\frac{1}{x}. \tag{11}$$

Now, let us consider the zeta function $\varsigma(2)$

$$\varsigma(2) = \sum_{x=1}^{\infty}\frac{1}{x^2} = \frac{\pi^2}{6} \tag{12}$$

and the harmonic series $H_n$ for a given value of $n$

$$H_n = \sum_{x=1}^{n}\frac{1}{x}. \tag{13}$$

It is known that

$$\lim_{n\to\infty} H_n - \ln(n) = \gamma, \tag{14}$$

where $\gamma = 0.5772156649\ldots$ is the Euler-Mascheroni constant. We can thereby approximate $\Lambda(P_n)$ for large $n$ as

$$\Lambda(P_n) \approx \frac{2\pi^2(n-2)}{3} - 4[\ln(n) + \gamma]. \tag{15}$$

Following similar calculation, we can obtain the values of $\kappa(C_n)$ for large $n$ as

$$\Lambda(C_n) = \begin{cases} \dfrac{2n\pi^2}{3} + \dfrac{8}{n} & \text{for } n \text{ even,} \\ \dfrac{2n\pi^2}{3} & \text{for } n \text{ odd.} \end{cases} \tag{16}$$

It is straightforward to realize that $\Lambda(S_n) \approx n^2$ and $\Lambda(K_n) \approx n^4$ for large $n$. For very large values of $n$, we have $\Lambda(P_n) \approx 2\pi^2(n-2)/3$ and $\Lambda(C_n) \approx 2\pi^2 n/3$. Because

$$\frac{2\pi^2(n-2)}{3} < \frac{2\pi^2 n}{3}, \text{ we then have } \frac{2\pi^2(n-2)}{3} - 4[\ln(n) + \gamma] < \frac{2\pi^2 n}{3}, \text{ which}$$

immediately implies that $\Lambda(P_n) < \Lambda(C_n)$. We thus have

$\Lambda(K_n) > \Lambda(S_n) > \Lambda(C_n) > \Lambda(P_n)$. This order follows our intuition; in the complete graph every vertex has the maximal possible degree and every pair of vertices are connected. The star graph, which is a subgraph of the complete graph, keeps one vertex with the maximal possible degree and all non-connected nodes are separated by only two links from each other. Finally, the path graph appears intuitively as the least clumped structure due to the low degree of its nodes (only one and two) and because of the large separation among them.

### 3.3. Bounds for the clumpiness coefficient

Following the line of the previous subsection, we can obtain the general bounds for the clumpiness coefficient. First, we can prove the following:

**Lemma 1**. Let $G = (V, E)$ be a connected graph having $n$ nodes. Then for any edge $e \in E$, we have

$$\Lambda(G - e) \leq \Lambda(G) \tag{17}$$

**Proof**. The result immediately follows from the following observations. For any node $i \in V$, we have that $k_i^G > k_i^{G-e}$, where $k_i^G$ is the degree of the node $i$ in the graph $G$ and $k_i^{G-e}$ is the degree of the node $i$ in the graph $G - e$. Moreover, for any pair of vertices $i, j \in V$ we have that $d_{ij}^{G-e} \geq d_{ij}^G$, where $d_{ij}^{G-e}$ and $d_{ij}^G$ are the topological distances of the vertices $i$ and $j$ in the graphs $G - e$ and $G$ respectively. Hence, we have

$$\Lambda(G) = \sum_{ij}^n \frac{k_i^G k_j^G}{\left(d_{ij}^G\right)^2} \geq \sum_{ij}^n \frac{k_i^{G-e} k_j^{G-e}}{\left(d_{ij}^G\right)^2} \geq \sum_{ij}^n \frac{k_i^{G-e} k_j^{G-e}}{\left(d_{ij}^{G-e}\right)^2} = \Lambda(G - e). \tag{18}$$

**Corollary 1**. Let $G = (V, E_1)$ and $H = (V, E_2)$ be two connected graphs on $n$ vertices such that $E_1 \subseteq E_2$, then we have $\Lambda(G) \leq \Lambda(H)$. In particular, we have $\Lambda(G) \leq \Lambda(K_n)$.

A graph is said to be Hamiltonian if there is a cycle, *i.e.,* a closed loop, which visits each node of the graph exactly once.

**Theorem 1**. Let $G$ be a connected graph having $n > 2$ nodes. Then,

a) if $G$ is Hamiltonian then $\Lambda(P_n) \leq \Lambda(G) \leq \Lambda(K_n)$,

b) $\Lambda(G) \geq \Lambda(S_{\Delta(G)})$, where $\Delta(G)$ is the maximum degree of the nodes of $G$.

**Proof**.

a) Since $G$ is Hamiltonian, $P_n$ is a subgraph of $G$. The result immediately follows from Corollary 1.

b) Clearly $S_{\Delta(G)}$ is a subgraph of $G$. Thus the result is a consequence of Corollary 1.

**Conjecture**. Let $T$ be any tree and $P_n$ and $S_n$ the path graph and the star graph on $n$ vertices, respectively. Then we have

$$\Lambda(P_n) \leq \Lambda(T) \leq \Lambda(S_n) \tag{19}$$

### 3.4. Relative clumpiness coefficient and classification of complex networks

In this section we are interested in proposing a method of selecting a cutoff value for the clumpiness parameter $\Lambda$ of a graph in order to determine whether the graph is clumped or not. Let us consider a graph $G$ having $n$ nodes and $m$ links. We have already proved that the maximum value of $\Lambda$ for a graph with $n$ nodes is $\Lambda(K_n) = n(n-1)^3/2$. However, for our $n,m$-graph this means to create new links up to $m = n(n-1)/2$. Instead we can think about the maximum value of $\Lambda$ that can be obtained for a graph having $m$ links. This is equivalent to rewiring the links of the $n,m$-graph to obtain the maximum clumpiness. The simplest way of doing that is to create the largest possible complete graph having $m$ links. In other words, we can divide the $n,m$-graph into a complete graph $K_{n_1}$ having $m$ links and $n_2$ isolated nodes with $n = n_1 + n_2$. With this intuition in mind we have that $n_1 = \dfrac{1+\sqrt{1+8m}}{2}$, which can be very well approximated to $n_1 \approx 0.5 + \sqrt{2m}$ for large $m$.

Then, the maximum clumpiness that can be obtained by rewiring an $n,m$-graph is $\Lambda(K_{n_1}) = n_1(n_1 - 1)^3/2$. Consequently, if we normalize the clumpiness coefficient of the $n,m$-graph by dividing it by $\Lambda(K_{n_1})$ we obtain the relative clumpiness coefficient $\Phi(G)$, which is defined and bounded as

$$0 \leq \Phi(G) = \dfrac{2\Lambda(G)}{\sqrt{n_1}(\sqrt{n_1}-1)^3} \leq 1. \qquad (20)$$

The upper bound is obtained when the graph has $m = n(n-1)/2$ links, i.e., for $K_n$. The lower bound is reached for very large graphs, $n \to \infty$. As we have already shown, the minimum value of $\Lambda$ is obtained for $P_n$, which makes that $\Phi(P_n) \to 0$ as $n \to \infty$.

The value of $\Phi(G)$ represents how clumped the graph $G$ is in relation to the most clumped graph that can be created by rewiring its links. Then, we can consider three classes of graphs: loose, clumped and very clumped. We consider that the graphs having less than 1/3 of the clumpiness of $K_{n_1}$ are loose, i.e., $\Phi(G) \leq 0.33$, those having $0.33 < \Phi(G) \leq 0.66$ are clumped and those having $\Phi(G) > 0.66$ are very clumped. Then, we consider that any network having $\Phi(G) > 0.33$ are clumped and those having $\Phi(G) \leq 0.33$ are loose.

### 3.5. Universality classes of complex networks

Here we analyze hypothetical networks having different topological organization of the most central nodes. We refer only to the degree centrality but the extension to any other centrality measure is straightforward. In this context, we consider four universality classes of complex networks illustrated in Fig. 4.

**Insert Fig. 4 about here.**

As we can see in Fig. 4a, one of these classes of networks is the one in which most central nodes are close to each other forming a clumped network. The mixing pattern of such networks consists of a series of highly connected nodes preferentially attached to each other while the less connected nodes are preferentially attached to other nodes with low connectivity. This mixing pattern is known as *assortative mixing*, that is "a preference for high-degree vertices to attach to other high-degree vertices" [22]. In this particular case we deal with *clumped assortative* networks. The clumped assortativity refers to the combination of an assortative mixing and a large clumpiness of the high-degree nodes. These networks must display large topological homogeneity, probably showing good expansion characteristics, *i.e.,* they do not contain structural bottlenecks [31, 32].

If the most connected nodes of the network are preferentially attached to nodes of low connectivity but keep a small distance among them, the network displays a *clumped disassortative* architecture (Fig. 4b). The disassortative mixing refers to the pattern where "high-degree vertices are attached to low-degree ones" [22]. The clumped disassortativity is then the combination of a disassortative mixing and a large clumpiness of the most connected nodes. This could appear counterintuitive at first sight, but it is typical, for instance, of complete bipartite (or almost bipartite) graphs, in which a few high-degree nodes are linked to each other over only one step of a large number of low-degree nodes. This connectivity pattern produces the disassortative mixing of the network and the small distance (only two steps separate a high-degree node from another) between the high-degree nodes gives its clumped nature.

On the other hand, the high-degree nodes in the network can be separated from each other by relatively large distances forming a class of not clumped, or loose networks (Fig. 4c). If these high-degree nodes are preferentially attached to each other leaving the least connected nodes to be directly interconnected, the network displays assortative mixing. The mixing pattern of this network represents a type of *loose assortative* organization. A typical organization of these networks is the formation of communities in which every community displays assortative mixing pattern. This makes the network as a whole display such assortative mixing. However, the separation of the high-degree nodes in one community from the high-degree nodes in another makes the clumpiness of the network decrease significantly. This makes the network display a loose mixing pattern. The community structure in complex networks has been shown to play a significant role in the dynamic processes taking place on the networks [33, 34].

The fourth organizational type of networks is formed by the class of *loose disassortative* networks (Fig. 4d). In these networks the high-degree nodes are preferentially attached to low-degree nodes, which makes the network displays disassortative mixing. In addition, the high-degree nodes are separated from each other by a relatively large number of links, which produces a significant decrease of the clumpiness.

**3.6. Generalization of the clumpiness coefficient**

We now mention a possibility of generalizing the clumpiness coefficient to ones based on other graph parameters. There are several centrality measures that have been defined and applied for the study of complex networks. In general, the notion of centrality comes from its use in social networks [17]. Intuitively, it is related to the ability of a node to communicate directly with other nodes, or to its closeness to many other nodes or to the quantity of pairs of nodes which need a specific node as intermediary in their communications [20]. Among well-known centrality measures, we can mention the betweenness or load centrality, the closeness centrality and the eigenvector centrality [17]. Other measures such as the subgraph centrality [21] have been recently proposed in the literature.

The clumpiness coefficient can be generalized for any centrality measure. It is defined as the averaged value of the product of the centrality measure for all pairs of nodes $C_i C_j$ in the network divided by a power of the corresponding topological distance $d_{ij}$ separating them:

$$\Lambda(G,C,\alpha) = \sum_{i<j}^{N} \frac{C_i C_j}{(d_{ij})^\alpha}. \tag{21}$$

As can be seen from this expression, when the most central nodes are directly connected, $d_{ij} = 1$, the clumpiness reaches its maximum. When the most central nodes are far away from each other, $d_{ij} \gg 1$, on the other hand, the clumpiness reaches its minimum. If **c** is a column vector of the centrality measure, we have

$$\Lambda(G,C,\alpha) = \frac{1}{2}(\mathbf{c}^T \mathbf{R} \mathbf{c}). \tag{21}$$

## 4. Spectral measure of clumpiness

### 4.1. Definition of the spectral measure of clumpiness

In addition to the clumpiness coefficient defined in Section 3.1, we also propose a spectral measure of clumpiness based on the clumpiness matrix $\hat{\mathbf{I}}$ defined in (5). Let $\{\varepsilon_1, \varepsilon_2, \cdots, \varepsilon_n\}$ be the nondecreasing order of the eigenvalues of $\hat{\mathbf{I}}$. We propose to use the principal eigenvalue of the clumpiness matrix (5) as a spectral measure of clumpiness:

$$\eta(G) = \varepsilon_1. \tag{22}$$

As we emphasized in Section 2, this is a measure of clumpiness with a different discriminant power.

The interpretation of this measure as a clumpiness index for a graph is given as follows. First of all we consider that the clumpiness index of a network is an additive function of node clumpiness,

$$\Lambda = \sum_{i=1}^{n} \Lambda_i ,$$

where $\Lambda_i$ represents the contribution of the node $i$ to the global clumpiness, which will be defined quantitatively later on. Now, let us consider a graph whose nodes can be ordered in nondecreasing order of clumpiness $\Lambda_1 \geq \Lambda_2 \geq \cdots \geq \Lambda_n$. We can thereby form a cluster by locating the most clumped node(s) at the centre, then the second most clumped node(s), then the thirds and so forth. The clumpiness of a network can be understood as a measure of the cohesiveness of the nodes in this "clumpiness cluster". In order to prove this meaning we first measure the participation of a node in the cluster by means of a column vector $\mathbf{x}$, whose $r$th entry captures the relative departure of the node $r$ from the centre $o$ of the cluster. The entries of the vector $\mathbf{x}$, $x_i$, take any values between zero and one. A value of zero corresponds to a node which is separated by an infinite distance from the centre of the cluster. In other words, $x_i = 0$ indicates that the node $i$ displays a very low clumpiness in contrast with nodes which are close to $o$. We impose the restriction that the norm of this vector $\mathbf{x}$ be one, $\mathbf{xx}^T = 1$.

Now, let us define a measure for the cohesiveness of the clumpiness cluster, $\eta$. A large cohesiveness of the nodes in this cluster indicates that most of the nodes are close to the centre $o$, or in other words that most of the nodes display large clumpiness. If the cohesiveness of the nodes in the cluster is low, it indicates that the graph displays low clumpiness. Let us now define formally the cohesiveness measure $\eta$. We can define this measure for the cohesiveness of the cluster in a similar way as in spectral clustering techniques [29]:

$$\eta = \sum_{i=1}^{n}\sum_{j=1}^{n} w_{ij} x_i x_j = \mathbf{x}^T \hat{\mathbf{I}} \mathbf{x} , \tag{23}$$

where $w_{ij} = \dfrac{k_i k_j}{d_{ij}^{\alpha}}$ is a weight assigned to every pair of nodes $(i, j)$ in the graph. The function $\eta$ increases with the increase of the clumpiness of the nodes as well as with the closeness of the nodes to the centre of the cluster. A maximally cohesive cluster can be found by maximizing the expression (23), which according to the Rayleigh-Ritz theorem [30] is given by

$$\eta = \max_{x}\left(\mathbf{x}^T \hat{\mathbf{I}} \mathbf{x} \,\middle|\, \mathbf{x}^T \mathbf{x} = 1\right) = \varepsilon_1, \tag{24}$$

where $\varepsilon_1$ is the spectral radius, the largest eigenvalue, of $\hat{\mathbf{I}}$, which is nothing but Eq. (22). Then the spectral measure (22) measures the cohesiveness of the nodes in the clumpiness cluster, and consequently, it represents a spectral measure of clumpiness. According to the Rayleigh-Ritz theorem [30], the optimal value of the participation vector is $\mathbf{x} = \mathbf{x}_1$, where $\mathbf{x}_1$ is the eigenvector corresponding to $\varepsilon_1$.

The sum of a row or column of the clumpiness matrix $\hat{\mathbf{I}}$ can be understood as the clumpiness of the corresponding node:

$$\Lambda_i = \sum_j w_{ij} \tag{25}$$

Then the interpretation of the principal eigenvector of $\hat{\mathbf{I}}$ as a relative participation of a node in the clumpiness cluster can be understood by means of the following analysis. The principal eigenvector of the matrix $\hat{\mathbf{I}}$ is proportional to the row sum of a matrix $\mathbf{M}$ formed by summing all powers of the clumpiness matrix, weighted by the corresponding powers of the reciprocal of the principal eigenvalue:

$$\mathbf{M} = \lim_{n \to \infty}\left\{\frac{1}{n}\left(\hat{\mathbf{I}} + \varepsilon_1^{-1}\hat{\mathbf{I}}^2 + \varepsilon_1^{-2}\hat{\mathbf{I}}^3 + \cdots + \varepsilon_1^{-n}\hat{\mathbf{I}}^{n+1}\right)\right\} \tag{26}$$

Let us consider a graph formed by three nodes having the following order of node clumpiness $\Lambda_1 > \Lambda_2 > \Lambda_3$. Then, we have that the sum of the rows of the matrix $\mathbf{M}$ follows the same order, $M_1 > M_2 > M_3$. Owing to the previously mentioned proportionality between the row sum of the matrix **M** and the principal eigenvector of $\hat{\mathbf{I}}$, we have $\mathbf{x}_1(1) > \mathbf{x}_1(2) > \mathbf{x}_1(3)$. Using our approach for building the clumpiness cluster, the first node is located at the centre, then the node 2 and finally the node 3. Then, the value of $\mathbf{x}_1(i)$, which is proportional to the closeness of the node $i$ to the centre of the cluster, measures the relative membership of such node to the cluster.

### 4.2. Statistical mechanical interpretation of the spectral measure of clumpiness

We here give a physical realization of the clumpiness matrix $\hat{\mathbf{I}}$. This enables us to give a statistical mechanical interpretation of the spectral measure of clumpiness $\eta(G)$.

We consider the tight-binding model, in which a particle moves among the nodes of a network. We assume that the hopping of a particle from one node to another is directly proportional to the degrees of the corresponding nodes. The physical intuition for this is as follow. We are considering connected networks. There is therefore always a path from one node to another. If the start node has degree $k_i$, there will be $k_i$ ways for the particle to leave the node. At the same time, if the goal node has degree $k_j$, the particle can arrive at it through $k_j$ different paths. We might then consider that the number of paths that the particle can follow from a node to another is proportional to the degrees of the two nodes. On another account, we can consider that the hopping is inversely proportional to the length of the path connecting both nodes. In short, we can make the hopping proportional to $\dfrac{k_i k_j}{(d_{ij})^\alpha}$, which is equivalent to saying that we consider the following tight-binding Hamiltonian:

$$H = -t \sum_{i,j} \frac{k_i k_j}{(d_{ij})^\alpha} (|i\rangle\langle j| + |j\rangle\langle i|) + \sum_i V_i |i\rangle\langle i|. \qquad (27)$$

For simplicity, we hereafter make $V_{ii} = V$ for every node of the network and we immediately obtain that the Hamiltonian is equal to $\mathbf{H} = V\mathbf{I} - t\hat{\mathbf{I}}$, where $\mathbf{I}$ is the identity matrix of order $n$, and $\hat{\mathbf{I}}$ is defined in (6):

$$\mathbf{H} = \begin{pmatrix} V & -t\dfrac{k_1 k_2}{(d_{12})^\alpha} & -t\dfrac{k_1 k_3}{(d_{13})^\alpha} & \cdots & -t\dfrac{k_1 k_n}{(d_{1n})^\alpha} \\ -t\dfrac{k_2 k_1}{(d_{21})^\alpha} & V & -t\dfrac{k_2 k_3}{(d_{23})^\alpha} & \cdots & -t\dfrac{k_2 k_n}{(d_{2n})^\alpha} \\ -t\dfrac{k_3 k_1}{(d_{31})^\alpha} & -t\dfrac{k_3 k_2}{(d_{32})^\alpha} & V & \cdots & -t\dfrac{k_3 k_n}{(d_{3n})^\alpha} \\ \vdots & \vdots & \vdots & \ddots & \vdots \\ -t\dfrac{k_n k_1}{(d_{n1})^\alpha} & -t\dfrac{k_n k_2}{(d_{n2})^\alpha} & -t\dfrac{k_n k_3}{(d_{n3})^\alpha} & \cdots & V \end{pmatrix} \qquad (28)$$

From now on, we set the origin of the energy scale to $V = 0$ and the unit of the energy scale to $t = 1$. We then use the Schrödinger equation for calculating the energy associated with the clumpiness of central nodes in a complex network:

$$\mathbf{H}|\psi_j\rangle = E_j|\psi_j\rangle, \qquad (29)$$

where $E_j$ and $\psi_j$ are the eigenvalues and eigenvectors of the $\mathbf{H}$ matrix, respectively. It is evident that $E_j = -\varepsilon_j$, where $\varepsilon_j$ are the eigenvalues of $\hat{\mathbf{I}}$. Consequently, we can define a clumpiness partition function for the network

$$Z_C = \mathrm{Tr}\, e^{-\beta \mathbf{H}} = \mathrm{Tr}\, e^{\beta \hat{\mathbf{I}}} = \sum_{j=1}^{n} e^{\beta \varepsilon_j}. \qquad (30)$$

We thus take account of lower eigenvalues than Eq. (22) with less weights specified by $\beta$. Using the clumpiness partition function, we can define the clumpiness entropy of the network

$$S_C(G, \beta) = -k_B \sum_{j=1}^{n} \left[ p_j (\beta \varepsilon_j - \ln Z_C) \right], \qquad (31)$$

where $p_j$ is the probability that the system occupies a microstate of energy $\varepsilon_j$,

$$p_j = \frac{e^{\beta \varepsilon_j}}{Z_C}. \qquad (32)$$

Then, we can write down Eq. (31) in the following equivalent way:

$$S_C(G,\beta) = -k_B \beta \sum_{j=1}^{n} \varepsilon_j p_j + k_B \ln Z_C \sum_{j} p_j, \qquad (33)$$

which, by using the standard relation $F = H - TS$, suggests the expressions for the clumpiness enthalpy and free energy of the network:

$$H_C(G,\beta) = -\frac{1}{Z_C} \sum_{j=1}^{n} \left( \varepsilon_j e^{\beta \varepsilon_j} \right), \qquad (34)$$

and

$$F_C(G,\beta) = -\beta^{-1} \ln Z_C. \qquad (35)$$

In the zero temperature limit, we have

$$Tr e^{\beta \hat{\mathbf{I}}} = \sum_{j=1}^{n} e^{\beta \varepsilon_j} \to e^{\beta \varepsilon_1} \text{ for large } \beta \text{ or as } T \to 0. \qquad (36)$$

Then, it is straightforward to realize that, in the same limit, the clumpiness enthalpy and free energy are equal to the negative of the spectral radius of $\hat{\mathbf{I}}$:

$$H_C(G, T \to 0) = F_C(G, T \to 0) = -\varepsilon_1. \qquad (37)$$

In other words, the spectral clumpiness coefficient, $\eta(G) = \varepsilon_1$, is the negative of the Gibbs free energy of the network in the zero temperature limit. In this limit, the network is "frozen" in the ground state which has the interaction energy $-\varepsilon_1$.

## 5. Numerical results

### 5.1. Artificial graphs

Our objective in this subsection is to study the general properties of the clumpiness coefficient and the statistical mechanical properties related to it in a series of small and simple graphs. With this objective in mind, we consider all possible 3-regular graphs (*i.e.* those graphs previously defined whose every node has degree 3) with 10 nodes. It is evident that for *k*-regular graphs, the clumpiness coefficient is given by

$$\langle \Lambda \rangle = k^2 \sum_{ij}^{n(n-1)/2} \frac{1}{(d_{ij})^2}$$ and we can study the specific influence of the separation of nodes to the graph clumpiness.

The average clumpiness coefficient for the 3-regular graphs studied varies from 3.8 to 4.5. In terms of the relative clumpiness coefficient $\Phi(G)$ in Eq. (20), this represent a change from 45.7 to 54%. According to the previous classification we have established all these graphs are clumped but no one is very clumped. The lowest value is obtained for the only one graph with the diameter (the maximal distance) equal to five (Fig. 5a), while the largest value is obtained for the Petersen graph (Fig 5b), in which every pair of non-connected nodes are separated by two links only. The average clumpiness coefficient $\langle \Lambda \rangle$ is poorly discriminant for these graphs. For instance, 5 non-isomorphic graphs are *degenerate*, having the same value of $\langle \Lambda \rangle = 4.222$; other three pairs of non-isomorphic graphs are also degenerate with identical values of the clumpiness coefficient, respectively. In short, the average clumpiness coefficient $\langle \Lambda \rangle$ is able to differentiate only 63% of the non-isomorphic 3-regular graphs studied.

**Insert Figure 5 about here.**

We next calculated the spectral measure of clumpiness, $\eta(G)$, which is equal to the negative Gibbs free energy in the zero temperature limit, $-\varepsilon_1$. The lowest and highest values are obtained for the same graphs as for the relative clumpiness coefficient $\Phi(G)$. In general, both magnitudes are strongly correlated with a correlation coefficient of 0.999, which is expected and desired because they are designed to measure the same network property. However, the spectral clumpiness coefficient is more discriminant than the relative clumpiness coefficient $\Phi(G)$ for this series of graphs. In fact, $\eta(G)$ discriminates 84% of the 3-regular graphs, showing identical values for a triple and a pair of non-isomorphic graphs only.

Finally, we calculated the statistical mechanical parameters, $S_C$, $H_C$ and $F_C$ for these 19 regular graphs. All the calculations are carried using $\beta = 0.1$. Remarkably, these functions discriminate 100% of the nonisomorphic regular graphs. In other words, there is not any single pair of graphs with identical values of these functions. The maximal entropy is reached for the graph having the lowest clumpiness, which is the graph having the lowest Helmholz and Gibbs free energies. On the other hand, the Petersen graph, which is the most clumped one, appears to be the least entropic 3-regular graph with 10 nodes. In general, there are nice correlations between the clumpiness coefficient and these statistical mechanical parameters.

In summary, the clumpiness coefficients as well as the statistical mechanical parameters changes regularly with the tiny changes in the structures of the graphs, which is a desired property for any graph theoretic descriptors. Based on our argument in Section 2 about the graph invariants and nonisomorphic graphs, we can say that the statistical mechanical parameters are more appropriate as clumpiness parameters than the single clumpiness coefficient, with their greater discriminant power.

### 5.2. Randomly evolved networks

In his seminal paper on assortative mixing in networks, Newman shows that for Erdős-Rényi (ER) random network, where links are placed at random regardless of the node degree, the assortativity coefficient is $R = 0$ in the limit of large graph size [22]. In addition, Newman also found that the Barabási-Albert (BA) model [13] shows no assortative mixing at all, showing that $R \to 0$ as $(\log^2 n)/n$ as $n$ becomes large [22]. Consequently, neither the ER nor the BA model reproduces the mixing patterns of networks and they are not able to reproduce any of the four universality classes found here.

We investigated how the relative clumpiness coefficient $\Phi(G)$ changes with the changes in the average degrees in these two models of random networks. In both models, each random network starts with *g* nodes and new nodes are added consecutively in such a way that a new node is connected to exactly *g* nodes chosen randomly from the already existing nodes. The average degree $\langle k \rangle$ is then exactly equal to $2g$. The new edges are attached according to a specific probability distribution, namely, the uniform distribution for the ER model and the preferential attachment mechanism for the BA model. We studied random networks grown by these two mechanisms up to $n = 1000$ nodes, changing systematically the value of $\langle k \rangle$ from 4 to 16. For every value of $\langle k \rangle$, we generated 100 random networks.

We found (Fig. 6) that the relative clumpiness coefficient $\Phi(G)$ of the networks generated by the ER model scales as a power-law of *g*, $\Phi_{ER} \sim \langle k \rangle^{\sigma}$, where $\sigma = 0.65$ (the correlation coefficient of the fitting is 0.995). All the ER networks obtained for $\langle k \rangle$ between 4 to 16 are loose, displaying low clumpiness. In order to obtain networks with large clumpiness with the ER model, we need values of $\langle k \rangle \geq 40$, which corresponds to very dense networks.

**Insert Figure 6 about here.**

On the other hand, the relative clumpiness coefficient $\Phi(G)$ of the BA networks scale as an exponential of $\langle k \rangle$, $\Phi_{BA} \sim \exp(\sigma \langle k \rangle)$, where $\sigma = 0.200$ (see Fig. 6). The correlation coefficient of the fitting is 0.997. Using this model, it is possible to generate clumped networks for values of $\langle k \rangle \geq 17.6$. A bigger difference is obtained when we try to generate very clumped networks. Using the ER model we need $\langle k \rangle \geq 115$, while by using the BA model a very clumped network can be obtained by using $\langle k \rangle \geq 25$. The open question then is how to generate loose networks with large average degree. Newman [22] has remarked that it "is an open question what type of network evolution processes could explain the values of *R* observed in the real-world networks". We also should take into account the clumpiness in considering this question.

The next question is to analyze how the statistical mechanics parameters change with the change of the clumpiness for randomly generated graphs. As a model parameter we selected the clumpiness entropy and analyze how it changes with the change of the relative clumpiness coefficient. In Fig. 7 we illustrate the plot of these two network parameters for graphs generated by using the ER and BA models having 1000 nodes. As can be seen both plots fit perfectly to a sigmoid function of the form

$$S(\beta = 0.001) = \frac{a}{1 + \exp(b\Phi + c)} \tag{39}$$

The correlation coefficient in both cases is larger than 0.99999, and the significance of the empirical parameters $a$, $b$ and $c$ will be evident further.

**Insert Fig. 7 about here.**

The plot in Fig. 7 clearly indicates that the clumpiness entropy of random networks change dramatically fast from its maximum to almost zero for a very narrow window of clumpiness values. For instance, for the case plotted in Fig. 7 the entropy changes from the maximum value $S = \ln(N)$ to almost zero by changing the relative clumpiness from 8% to 12%. Then, the parameter $a$ in (39) that controls the size of the sigmoid is evidently equal to $\ln(N)$.

In order to find the values of the parameters $b$ and $c$ we have generated the plots of the clumpiness entropy versus the relative clumpiness coefficient for different values of $N$. For the sake of brevity we study only the networks generated by using the ER model. We have obtained the sigmoid plots for ER networks having 29, 50, 150, 250, 500 and 1000 nodes. Then, by fitting we have observed that the parameters $b$ and $c$ scales as power-law of the number of nodes having correlation coefficients larger than 0.999,

$$b = c_1 N^\gamma \equiv 0.044723 N^{0.596142},$$

$$c = c_2 N^\delta \equiv 18.98589 N^{0.048962}.$$

By using these parameters in Eq. (39) we have generated the sigmoid functions for $N = 3000$, 2000, 100 and 15, which are plotted in Fig. 8 together with those previously obtained by fitting.

**Insert Fig. 8 about here.**

The dramatic decrease of the entropy with the increase of the relative clumpiness can be understood by considering the following facts. The largest entropy is obtained for a fully disconnected network in which every node has degree equal to zero and then $S = \ln(N)$. That is, in the fully-disconnected network every node is indistinguishable from each other. When we have a connected network we can group together all nodes according to their degrees. In a path, for instance, all nodes except two have degree 2 which makes then indistinguishable to each other and consequently the entropy is close to the maximum. Of course, the number of groups consisting of nodes with the same degree increases as the average degree of the network increases. As a consequence the number of distinguishable nodes (according to their degrees) also increases, which makes that the entropy decreases dramatically. This situation can be observed in Fig. 9, where we have plotted the normalized degrees for ER networks having different average degrees. In this figure we can observe that the number of groups of nodes with the same degree increases dramatically by changing the average degree from 3.98 to 11.71 and it is even larger for $\langle k \rangle = 22.85$.

**Insert Fig. 9 about here.**

Because the high plateau of the sigmoid function depends on the logarithm of the number of nodes, for small networks the range of entropy values is very much reduced in comparison to larger networks. Consequently, there is an "envelope" function that determines how the entropy of ER networks decreases with the increase of clumpiness.

In fact, this function determines the upper limit for which a network having a given clumpiness can increases its entropy. The envelope function is given by first fixing $\Phi$ and maximizing the entropy with respect to $N$. We first solve the equation

$$0 = \frac{\partial}{\partial N} \frac{\ln N}{1 + \exp(c_1 N^\gamma \Phi + c_2 N^\delta)}, \qquad (40)$$

which gives

$$1 + \exp(-c_1 N^\gamma \Phi - c_2 N^\delta) = (c_1 \gamma N^\gamma \Phi + c_2 \delta N^\delta) N \ln N. \qquad (41)$$

We find a numerical solution of this equation for each value of $\Phi$. The solution is then a function of $\Phi$, which we denote by $N(\Phi)$. We then input this in the first equation and have

$$\rho(\Phi) = \frac{\ln N(\Phi)}{1 + \exp\left(c_1 N(\Phi)^\gamma \Phi\right) + c_2 N(\Phi)^\delta}. \qquad (42)$$

This means that the entropy of a network generated by the ER model cannot takes values over $\rho(\Phi)$; $S(\beta = 0.001) \leq \rho(\Phi)$.

### 5.3. Real-world networks

Here we study 30 real-world networks representing social, informational, technological, biological and ecological systems. The social networks include a network of the corporate elite in the US [35], inmates in prison, injectable drug users (IDUs), the Zachary karate club, college students on a course about leadership, the friendship ties among 31 physicians (Galesburg) [363] and a sexual network in Colorado Springs [37]. The informational and technological networks include two semantic networks, one based on Roget's Thesaurus of English (Roget) and the other on the Online Dictionary of Library and Information Science (ODLIS). They also include three citation networks: one consisting of papers published in the *Proceedings of Graph Drawing* in the period 1994–2000 (GD), and papers published or citing articles from *Scientometrics* for the period 1978–2000 (SciMet), papers containing the phrase "Small World" [36]. The two technological networks are the airport transportation network in the US in 1997 [36] and the Internet at the autonomous systems (AS) level as from April 1997 [38]. The biological networks are the protein–protein interaction networks (PINs), for *Saccharomyces cerevisiae* (yeast) [39] and for the bacterium *Helicobacter pylori* [40]; two transcription interaction networks concerning *E. coli* and yeast [41]; and the neural network in *C. elegans* [12]. The protein residue networks correspond to the proteins with Protein Data Bank (PDB) codes: the immunoglobulin 1A4J; the serine protease inhibitor 1EAW and the oxidoreductase 1AOR. In these networks each residue is represented as a single node, centered on $C_\beta$ atoms. Then a contact map is represented by taking a 7 Å cutoff radius [42]. Finally, the ecological networks studied correspond to the following food webs [43]: Benguela, Bridge Brook, Coachella Valley, El Verde rainforest, Little Rock Lake, Scotch Broom, St. Marks Seagrass, and Stony.

We illustrate in Fig. 10 the plot of the assortativity coefficient versus the relative clumpiness coefficient $\Phi(G)$ expressed in percentage for the studied real-world networks. The assortativity coefficient $R$ is simply the Pearson correlation coefficient of the degrees at either ends of a link [22]. The negative values of $R$ indicate that the network is disassortative and the positive values that the network is assortative.

**Insert Fig. 10 about here.**

By simple inspection of Fig. 10 we can observe that the four classes of mixing patterns (clumped and loose assortative as well as clumped and loose disassortative) are represented in this selection of real-world networks. The most populated class of networks corresponds to the loose disassortative ($36.7\%$), which is followed by the clumped assortative ($30.0\%$). On the other side, the least populated class is the one of clumped assortative networks, which is represented only by three ecological networks. In general, there are more loose networks than clumped ones, i.e., 60% versus 40%, respectively.

The second important observation is that the classification of networks into these four classes is not determined by the type of functional systems that they represents, *e.g.*, social, biological, ecological, *etc*. However, it is important to observe that all biological networks (100%) are loose as well as most of social networks (71.4%). On the other hand, all ecological networks (100%) are clumped. In fact, the only two networks which are very clumped ($\Phi > 0.66$) are food webs.

In general, clumped networks display large average degree. The correlation coefficient between the average degree and $\Phi(G)$ for these 30 networks is 0.62. However, a large average degree does not guarantee that the network is clumped. As we have previously seen, a network with large number of high-degree nodes which are separated to each other by relatively large distances, displays loose characteristics despite that it has large average degree (see Fig 4c). For instance, the corporate elite network displays a large average degree $\langle k \rangle = 14.6$. However, the corporate elite network is a loose network having $\Phi(G) = 13.1$ because of large distances between the top elites. In a similar way the size of the complex network does not explain their clumpiness characteristics. We have seen that the only two very clumped networks are very small food webs having around 30 nodes. However, the correlation coefficient between the size and $\Phi(G)$ is only $-0.38$ for the 30 networks studied. In fact, among the networks having less than 500 nodes there are networks with values of $\Phi(G)$ ranging from 3 to 80.

Finally, we analyze the relationship between the relative clumpiness coefficient $\Phi(G)$ and the clumpiness entropy $S(G)$ for these real-world networks. In Fig. 11 we plot both parameters for these real-world networks, where we also plot the envelope function obtained previously for random networks.

**Insert Fig. 11 about here.**

As can be seen in Fig. 11 almost all real-world networks are located below the envelope function obtained for random networks. This means that, in general, for every real-world network exists an ER random graph having the same relative clumpiness coefficient $\Phi(G)$ but having the maximum possible clumpiness entropy $S(G)$. The only one exception is the sexual network of Colorado Spring, which has been previously identified as possessing other differential characteristics respect to the rest of real-world networks [44]. It is also very characteristic of this plot that most of the real-world networks are concentrated either close to the bottom line of the plot or to the envelope function. This means that there is a gap between the maximum and minimum possible entropies. At present we do not have a rational explanation for this observation.

## 6. Summary

In the present paper, we defined several measures of clumpiness, namely the clumpiness coefficient, the spectral measure of clumpiness and statistical mechanical quantities of clumpiness. We presented bounds of the clumpiness coefficient. We also present physical interpretations of the statistical mechanical quantities of clumpiness.

We then proposed to categorize complex networks into four classes with the use of the clumpiness and the assortativity. We demonstrated the classification, first for 3-regular graphs with 10 nodes, then for ER and BA random networks, and finally for real-world networks. This method successfully classifies 30 real-world networks into four classes of clumped assortative, clumped disassortative, loose assortative and loose disassorative networks. We also showed that the clumpiness coefficient successfully differentiated the ER model from the BA model; they could not be differentiated by the assortativity coefficient. We finally showed numerically a relation between the clumpiness coefficient and the clumpiness entropy for the ER random networks. The relation seems to hold for real-world networks as well.

**Acknowledgements**. E.E. thanks the Program "Ramón y Cajal", Spain for partial financial support.

**References**

[1] Nakanishi N, 1971 *Graph Theory and Feynman Integrals* (New York: Gordon & Breach)

[2] Chertkov M and Chernyak V Y, 2006 *J. Stat. Mech*. **P06009**

[3] Nicolai H, Peeters K and Zamaklar M, 2005 *Class. Quantum Grav*. **22** R193

[4] Gnotzmann S and Smilansky U, 2006 *Adv. Phys*. **55** 527

[5] Chakrabarti N K, Chakraborti A and Chatterjee A (eds), 2006 *Econophysics and Sociophysics: Trends and Perspectives* (Berlin: Wiley-VCH)

[6] Ben-Naim E and Redner S, 2005 *J. Stat. Mech*. **L11002**

[7] Strogatz S H, 2001 *Nature* **410** 268

[8] Albert R and Barabási A-L, 2002 *Rev. Mod. Phys*. **74** 47

[9] Bornholdt S and Schuster H G (eds), 2002 *Handbook of Graphs and Networks: From the Genome to the Internet* (Belin: Wiley)

[10] Newman M E J, 2003 *SIAM Rev* **45** 167

[11] Boccaletti S, Latora V, Moreno Y, Chávez M and Hwang D-U, 2006 *Phys. Rep*. **424** 175

[12] Watts D J and Strogatz S H, 1998 *Nature* **393** 440

[13] Barabási A L and Albert R, 1999 *Science* **286** 509

[14] Costa L da F, Rodríguez F A, Travieso G and Villa Boas P R, 2007 *Adv. Phys*. **56** 167

[15] Park J and Newman M E J, 2005 *J. Stat. Mech*. **P10014**

[16] Albert R, Jeong H and Barabási A-L, 2000 *Nature* **406** 378

[17] Wasserman S and Faust K, 1994 *Social Network Analysis* (Cambridge: Cambridge University Press)

[18] Jordán F, Liu W-C, Davis A J, 2006 *Oikos* **112** 535

[19] Estrada E, 2007 *Ecol. Complex*. **4** 48

[20] Borgatti S P and Everett M G, 2006 *Social Networks* **28** 466

[21] Estrada E and Rodríguez-Velázquez J A, 2005 *Phys. Rev*. **71** 056103

[22] Newman M E J, 2002 *Phys. Rev. Lett*. **89** 208701

[23] http://en.wikipedia.org/wiki/Graph_property

[24] Essam J W and Fisher M E, 1970 *Rev. Mod. Phys*. **42** 272

[25] Read R and Corneil D, 1977 J. Graph Theor. **1** 339

[26] Gorni G and Tutaj-Gasińska H, 2004 *Comm. Algebra* **32** 495

[27] Estrada E and Rodríguez L, 1997 *MATCH: Comm. Math. Comput. Chem*. **35** 157

[28] Harary F, 1969 *Graph Theory* (Reading, MA: Addison-Wesley)

[29] Sarkar S and Boyer K L, 1998 Comput. Vision Image Unders. **71** 110

[30] Horn R A and Johnson C R, 1990 *Matrix Analysis* (Cambridge: Cambridge Univ. Press)

[31] Donetti L, Neri F and Muñoz M A, 2006 *J. Stat. Mech*. **P08007**

[32] Estrada E, 2006 *Europhys. Lett*. **73** 649

[33] Lambiotte R and Ausloos M, 2007 *J. Stat. Mech.* **P08026**

[34] Huang W and Li C, 2007 *J. Stat. Mech.* **P01014**

[35] Davis G F, Yoo M and Baker W E, 2003 *Strategic Organization* **1** 301

[36] Batagelj V and Mrvar A, (http://vlado.fmf.uni-lj.si/pub/networks/data/)

[37] Potterat J J, Philips-Plummer L, Muth S Q, Rothenberg R B, Woodhouse D E, Maldonado-Long T S, Zimmerman H P and Muth J B, 2002 *Sex. Transm. Infect*. **78** i159

[38] COSIN database: http://www.cosin.org.

[39] Bu D, Zhao Y, Cai L, Xue H, Zhu X, Lu H, Zhang J, Sun S, Ling L, Zhang N, Li G and Chen R, 2003 *Nucleic Acids Res*. **31** 2443

[40] Rain J C, Selig L, De Reuse H, Battaglia V, Reverdy C, Simon S, Lenzen G, Petel F, Wojcik J, Schachter V, Chemana Y, Labigne A and Legrain P, 2001 *Nature* **409** 211

[41] Milo R, Itzkovitz S, Kashtan N, Levitt R, Shen-Orr S, Ayzenshtat I, Sheffer M and Alon U, 2004 Science **303 1538**

[42] Rana Atilgan A, Akan P and Baysal C, 2004 *Biophys. J.* **86** 85

[43] Dunne J A, Williams R J and Martinez N D, 2002 *Proc. Natl. Acad. Sci. USA* **99** 12917

[44] Estrada E, Hatano H, 2007 arXiv:0707.0756v1.

**Figure Captions**

**Figure 1.** Illustrative examples of two real-world networks with assortative mixing, where some of the high-degree nodes are clumped (a) or spread across the network (b). The assortativity coefficients ($r$) are displayed.

**Figure 2**. Illustrative examples of two real-world networks with disassortative mixing, where some of the high-degree nodes are clumped (a) or spread across the network (b). The assortativity coefficients ($r$) are displayed.

**Figure 3**. Plots of the normalized clumpiness coefficient of the 19 cubic regular graphs for the different values of $\alpha$.

**Figure 4**. Four classes of networks classified by the clumpiness and the assortativity.

**Figure 5.** a) The graph with the lowest value of the clumpiness coefficient among the 3-regular graphs having 10 nodes. b) The Petersen graph, which is the graph with the largest value of the clumpiness coefficient among the 3-regular graphs having 10 nodes.

**Figure 6**. A plot of the relative relative clumpiness coefficient $\Phi$ in percentage versus the average degree of networks, $k$ generated at random by using the ER (broken line) and BA (solid line) models. The dotted line represents the limit over which a network is considered as clumped.

**Figure 7**. Sigmoidal fits of the clumpiness entropy of random networks as a function of the relative clumpiness coefficient. The plot is the average of 100 realizations using the ER and BA models. We display the values of the average degree $k$ for some ER

networks in order to illustrate the trend followed in the plot by the change of this parameter.

**Figure 8**. Sigmoidal fits of the entropy of random networks as a function of the relative clumpiness coefficient for different network sizes. The network size decreases from left to right ($N$=3000, 2000, 1000, 500, 250, 150, 100, 50, 29, 15). The plots represented as solid lines are obtained by fitting both parameters using the data points represented as filled circles. The plots represented as dotted lines are those generated by using the Eq. (39) of the main text. The discontinuous line represents the envelope function obtained numerically (see text for explanations).

**Figure 9**. Plot of the normalized degrees of the nodes in the ER random networks with different average degrees. The nodes are ranked in decreasing order of their relative degrees.

**Figure 10.** Classification of real-world networks with the relative clumpiness coefficient and the Newman assortative coefficient. The symbols S, I, T, B and E denote social, informational, technological, biological and ecological networks, respectively; see the main text for details. The vertical solid and discontinuous lines represent the thresholds over which a network can be considered as clumped or very clumped, respectively.

**Figure 11**. Plot of the entropy versus relative clumpiness for real-world networks. The discontinuous curve represents the envelope function obtained for the ER networks.

**Figure 1**

a)

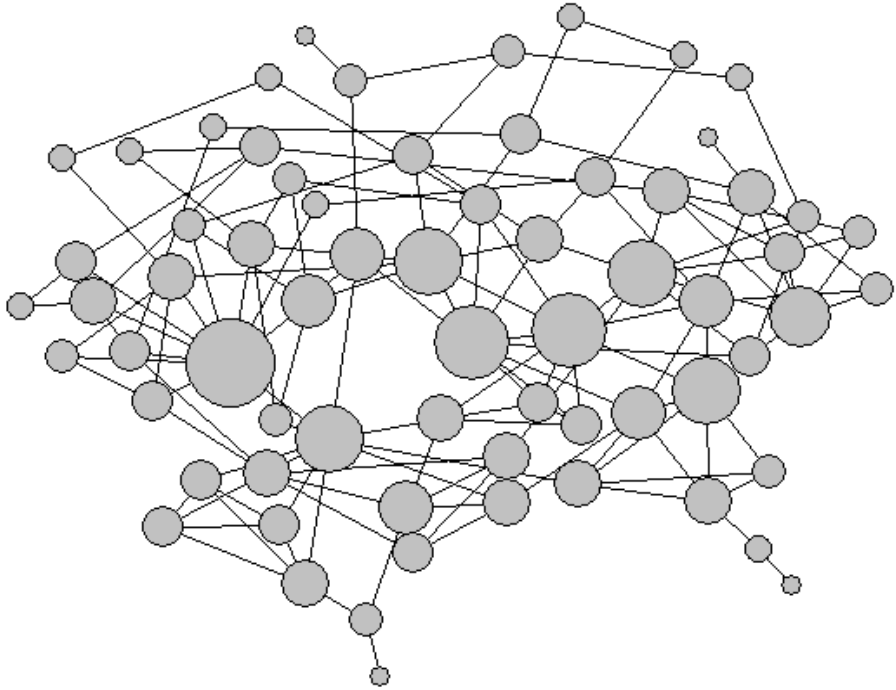

$r = 0.103$

b)

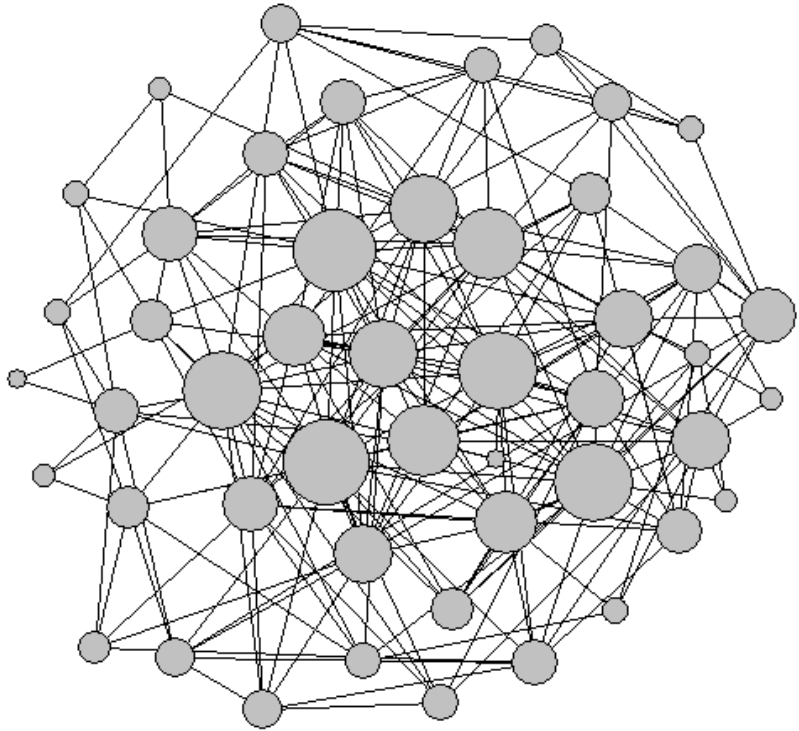

$r = 0.118$

**Figure 2**

a)

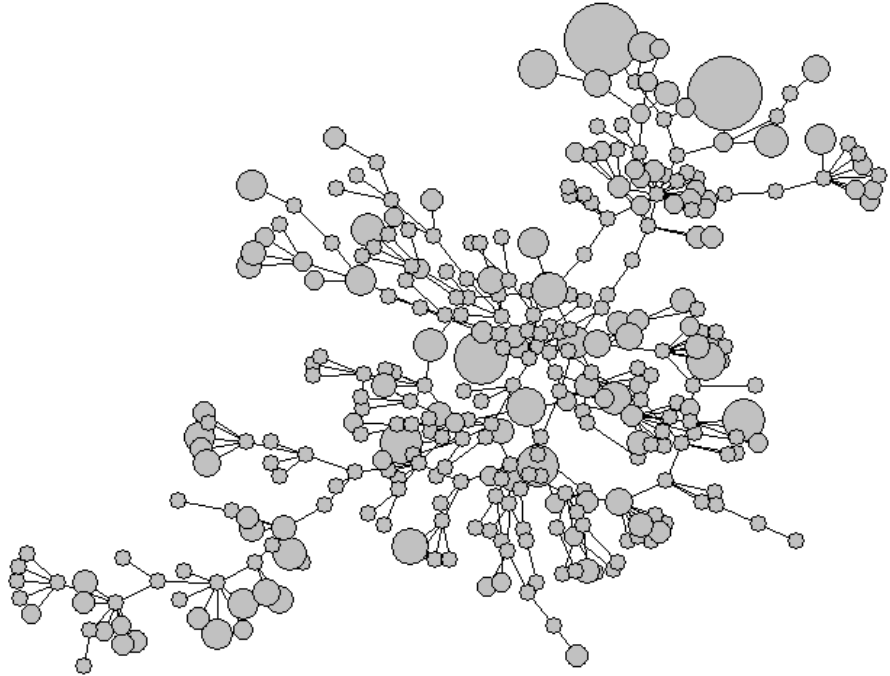

$r = -0.277$

b)

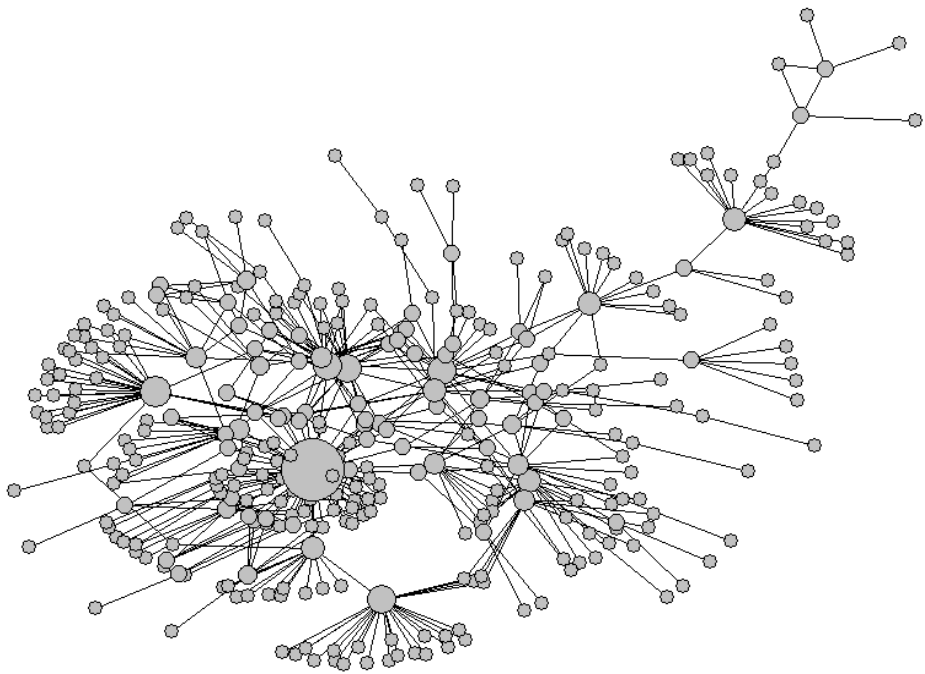

$r = -0.265$

**Figure 3**

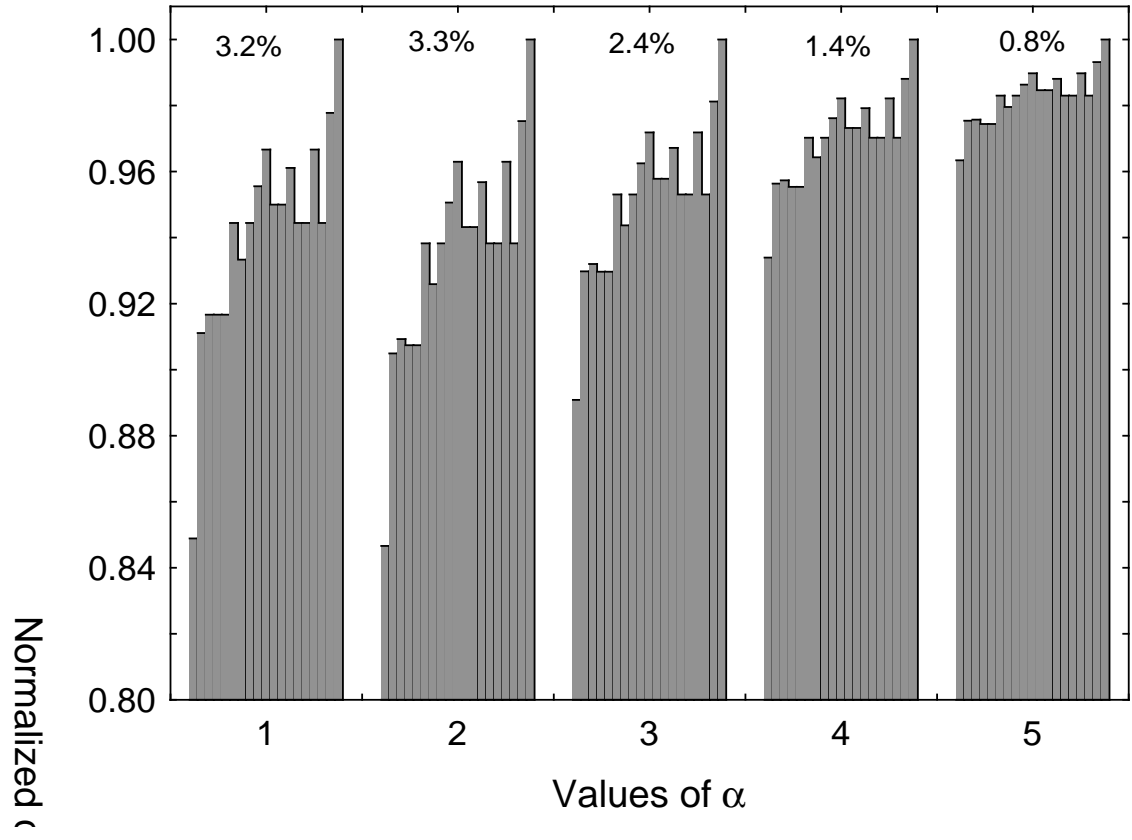

**Figure 4**

**a) Clumped Assortative**

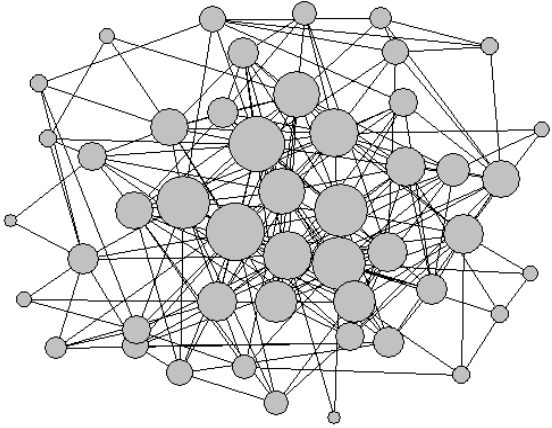

**b) Clumped Disassortative**

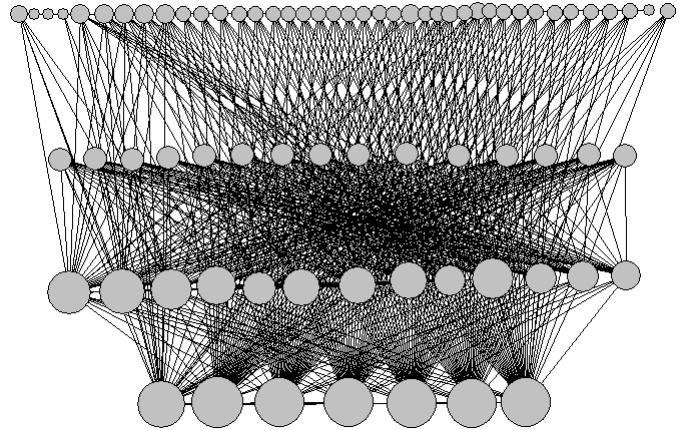

**c) Loose Assortative**

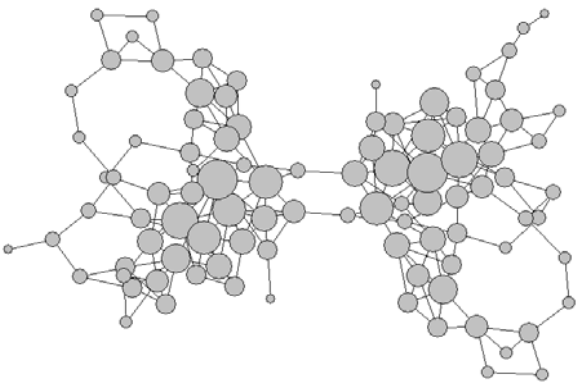

**d) Loose Disassortative**

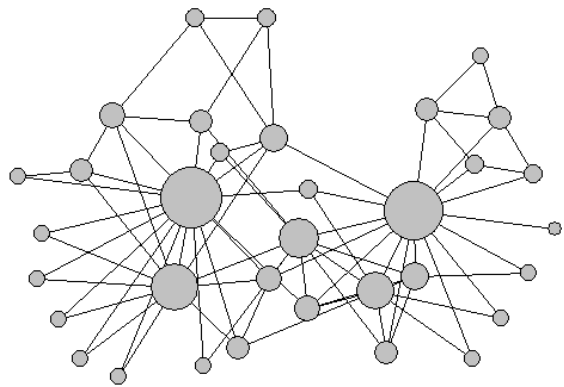

**Figure 5**

a) 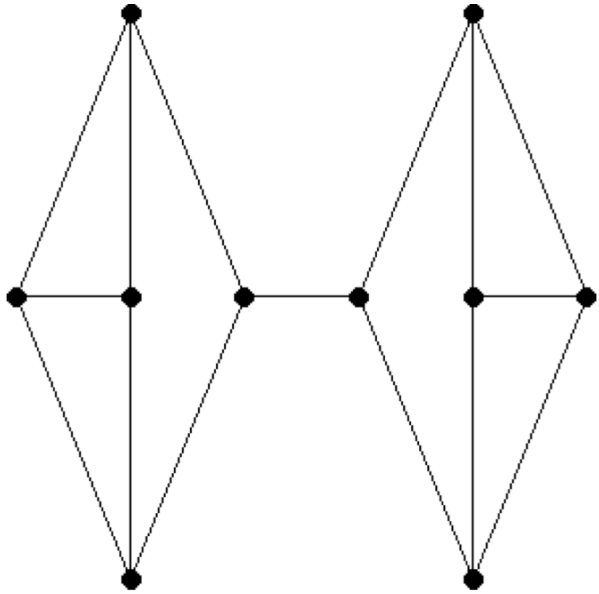    b) 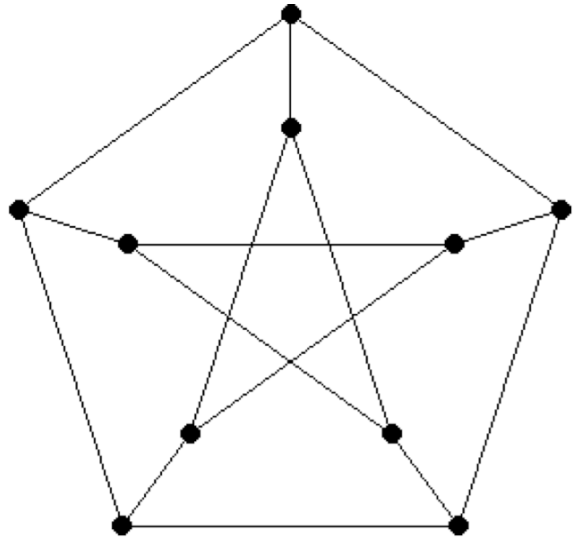

**Fig. 6**

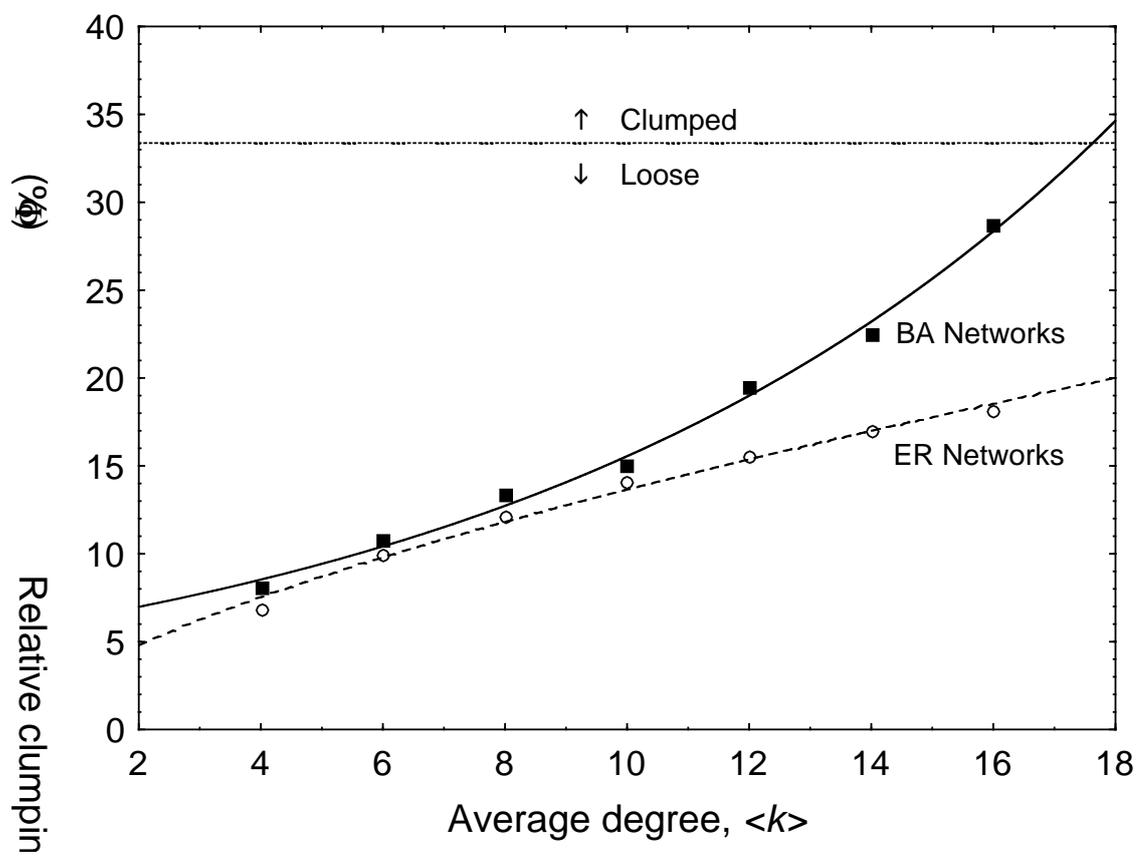

**Fig. 7**

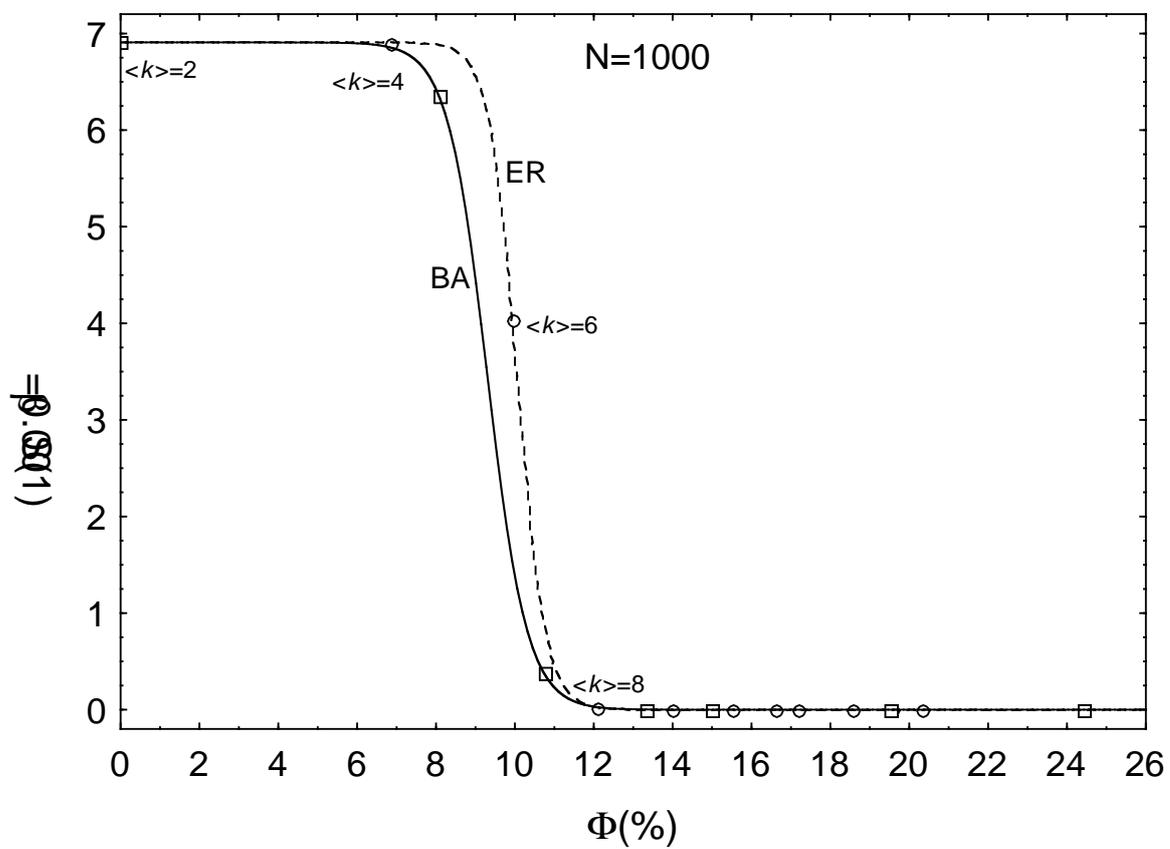

**Fig. 8**

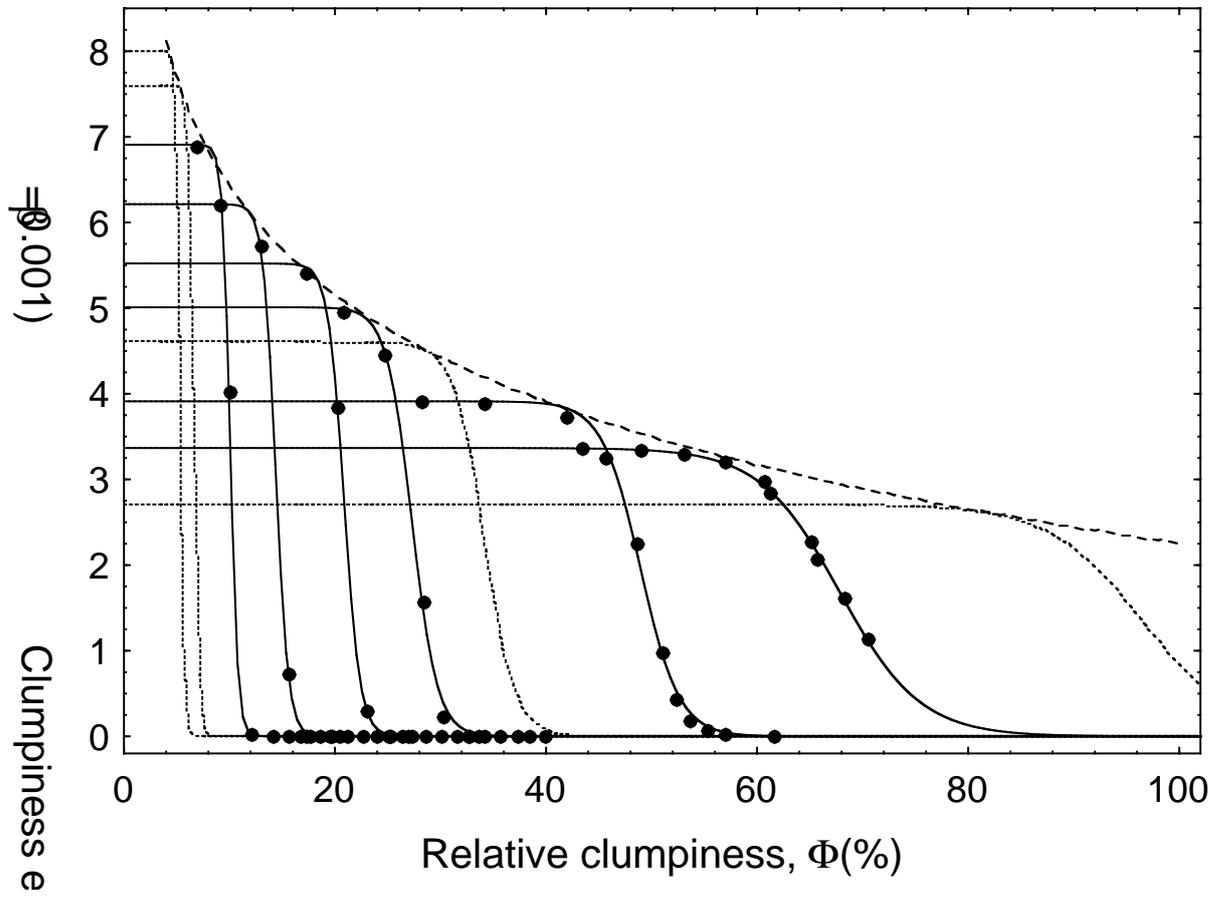

**Fig. 9**

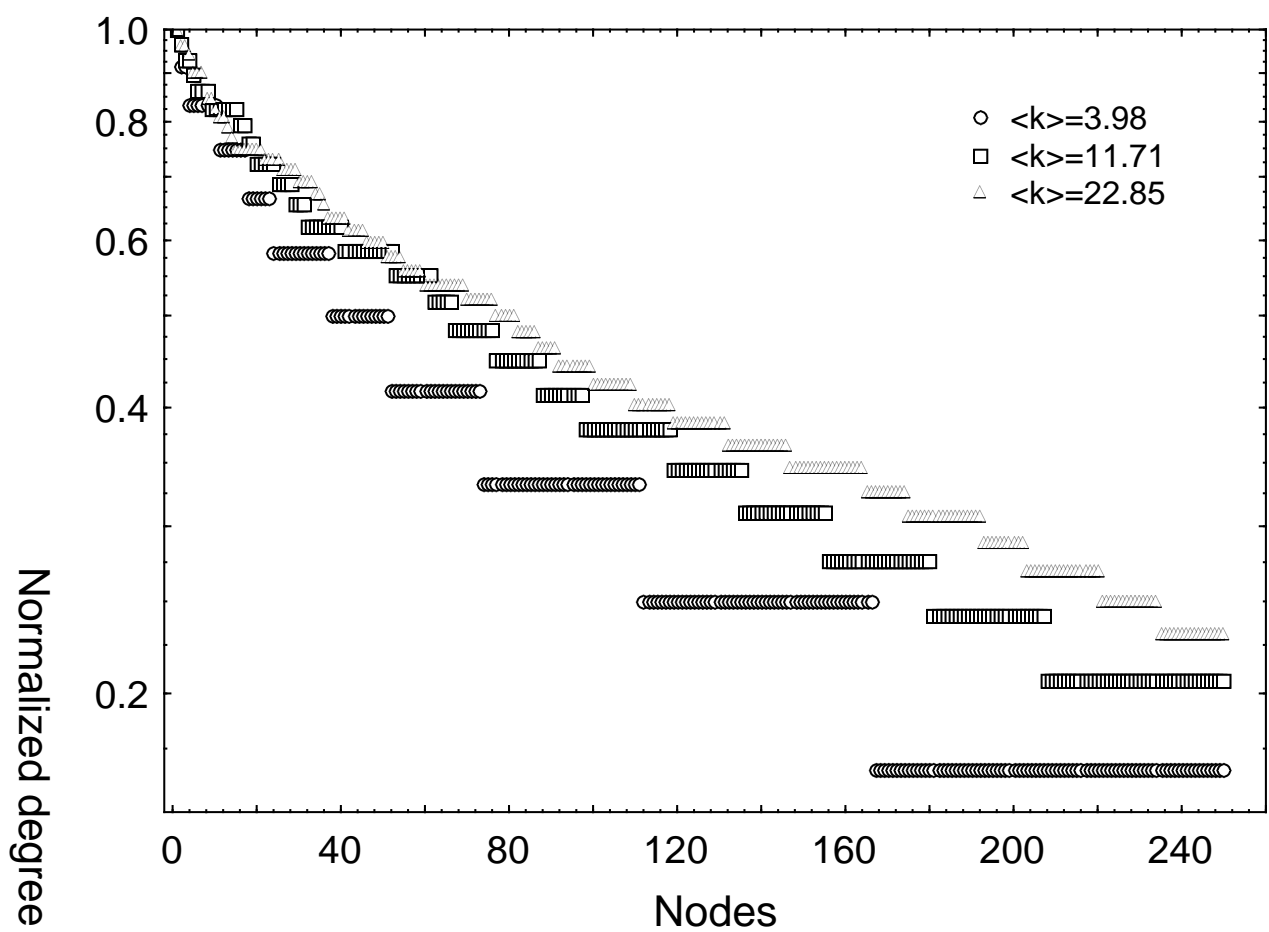

**Figure 10**

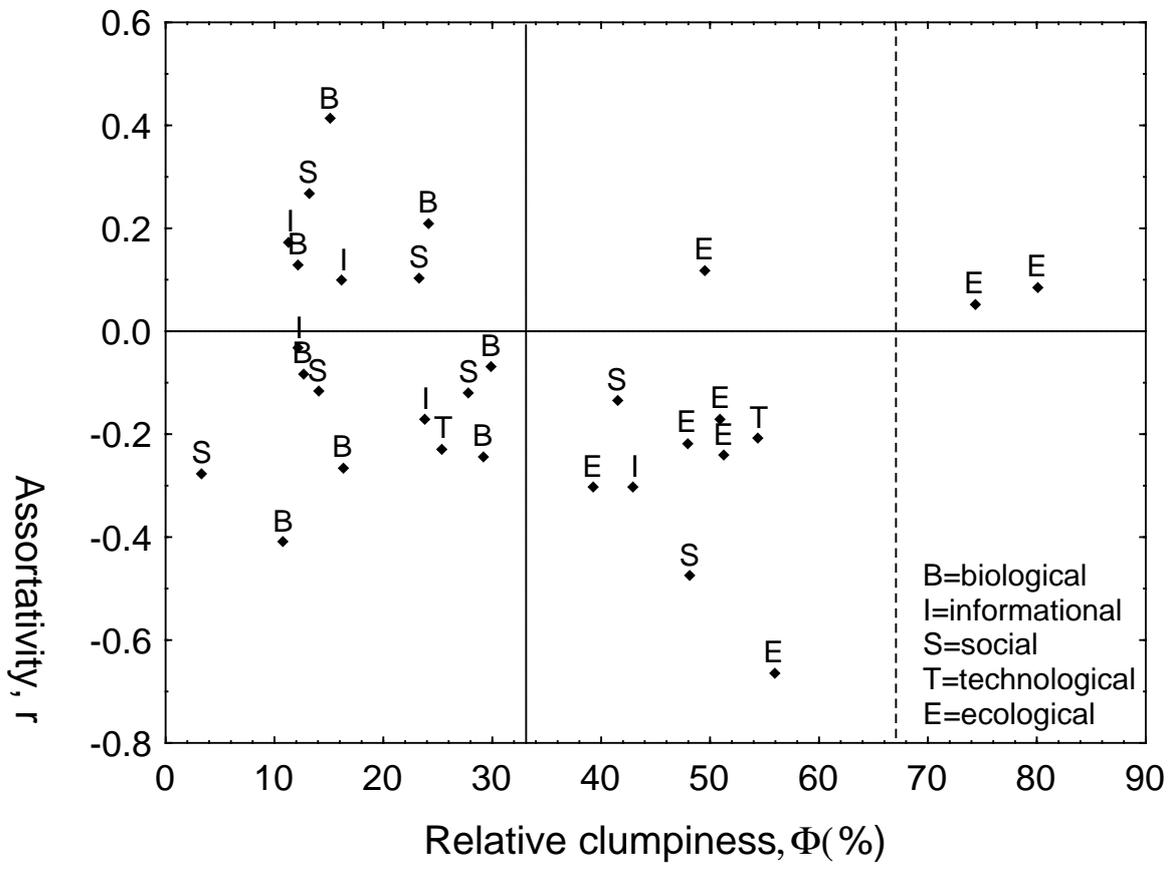

**Figure 11**

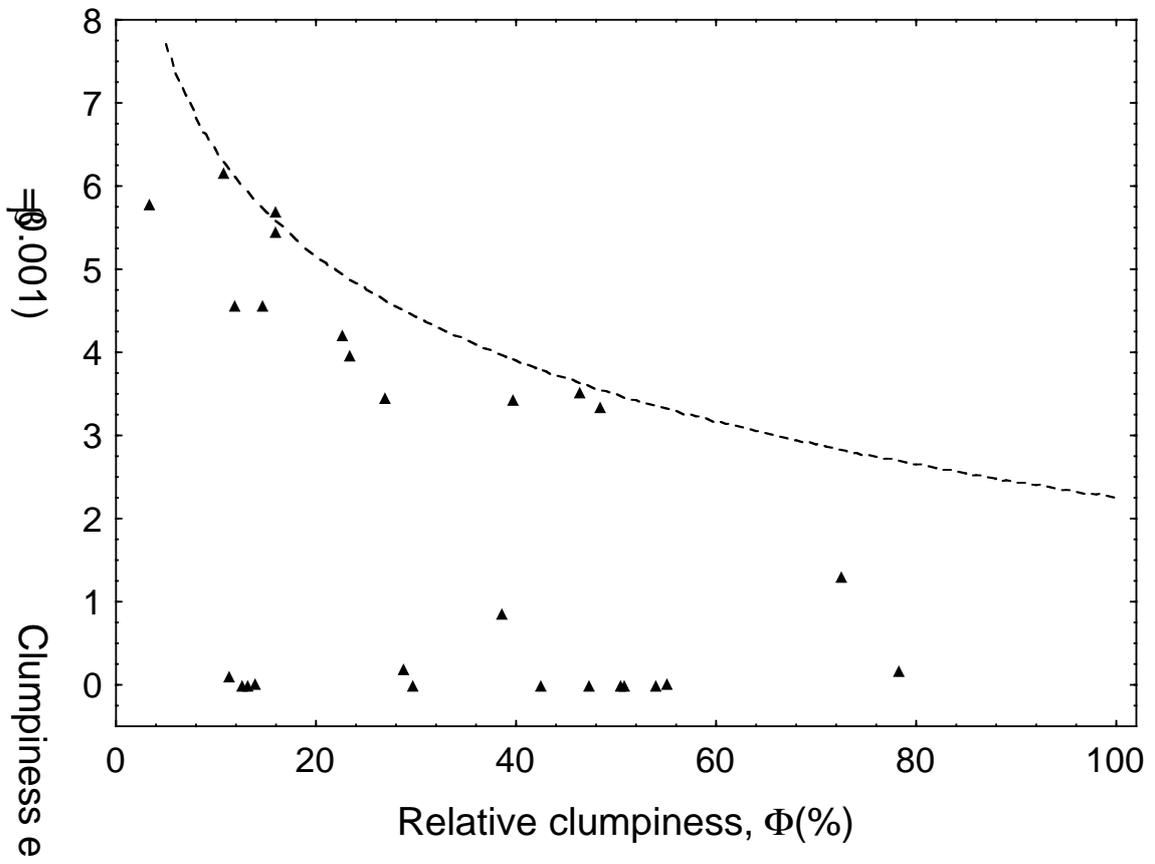